# Measuring ultrafast laser pulses using a single-shot amplitude swing implementation

CRISTIAN BARBERO,[1,*] ÍÑIGO J. SOLA [1,2], AND BENJAMÍN ALONSO[1,2]

[1]*Grupo de investigación en Aplicaciones del Láser y Fotónica (ALF), Universidad de Salamanca, 37008 Salamanca, Spain*
[2]*Unidad de Excelencia en Luz y Materia Estructuradas (LUMES), Universidad de Salamanca, Spain*
*cristianbp@usal.es*

**Abstract:** Single-shot characterization techniques are crucial when dealing with shot-to-shot pulse-shape fluctuations (e.g., unstable laser systems, high-power, or with low repetition rate) since the scanning configurations cannot measure single pulses. The demand for simple setups that can be easily adapted to a wide variety of experimental conditions is continuously rising. In this work, we propose a single-shot implementation of amplitude swing, maintaining the compactness, versatility, and robustness of the scanning versions of this technique. First, we theoretically study the proposed implementation, based on a pair of uniaxial wedges. Then, we present the retrieval ptychographic algorithm. Finally, we experimentally demonstrate the setup by comparing the single-shot and scanning traces and their retrieved pulses. In sum, we provide the ultrafast science community with a simple and versatile setup capable of measuring single laser pulses, which is necessary for characterizing fluctuating pulse trains, meeting the current increasing demand.

## 1. Introduction

The temporal characterization of ultrashort laser pulses is fundamental in their multiple applications, such as microprocessing of materials [1], femtochemistry and femtobiology [2], or strong field physics [3]. This characterization is typically done with self-referenced techniques, i.e., the pulse interacts with itself or with a replica of itself during the measurement. Nowadays, there are some well stablished techniques, that could be classified in spectrographic, as FROG [4], d-scan [5] or FROSt [6], and interferometric, as SPIDER [7] or self-referenced spectral interferometry (SI) [8]. Another approach is TIPTOE [9], based on measuring electric currents from sub-cycle tunneling ionization. In the spectrographic methods, the spectrum of a nonlinear signal is recorded while some parameters of the input pulse are scanned (e.g., delay between two replicas in FROG, or amount of dispersion in d-scan). This way, it is obtained a two-dimensional trace that encodes the pulse phase, retrieved in general with an iterative algorithm. For traces acquired in this way, the reconstructed pulse is actually an average of all the pulses collected during the scan. This is not a problem for high shot-to-shot stability systems, where all the emitted pulses are quite similar. Nevertheless, it prevents measuring single pulses, a required task when operating in single-pulse mode or dealing with shot-to-shot pulse-shape fluctuations. This is a common case for low repetition rate, high-power, and some fiber lasers. Interferometric techniques can solve this issue since they can be inherently single shot, provided enough energy to obtain the signal from a single laser shot, and that no additional shots are required for calibration. The pulse can be analytically reconstructed from a one-dimensional interferogram; however, they typically require complex setups, high stability, and precise alignment. Pulse instabilities have been theoretically studied by comparing FROG and SPIDER [10], as well as with d-scan [11]. Experimentally, FROG [12] and self-calibrating d-scan [13,14] have been used to measure pulse-shape instabilities in fiber lasers. The pulse train can be described by the so-called coherent artifact, the shortest repeatable pulse structure within the train (not necessarily the Fourier-limited pulse), and a much longer average pulse. This analysis quantifies the instabilities, but it still does not provide information about the shape of each individual pulse.

Another approach to measure single pulses is to use single-shot spectrographic techniques, where the varying parameter in the scanning implementation is translated into a spatial dimension. GRENOUILLE [15], a single-shot version of FROG, uses a Fresnel biprism and a thick nonlinear crystal to obtain the spatially resolved 2D FROG trace. The first single-shot implementations of d-scan [16,17] replace the moving pair of wedges used in the scanning setup by a prism, so the pulse propagates through different amounts of material along one transverse axis of the laser beam. Another single-shot d-scan [18] is based on transverse second harmonic generation (SHG), where dispersion is accumulated longitudinally along the nonlinear medium. In both schemes, the spatially resolved trace is recorded by an imaging spectrometer. There are also single-shot implementations of other techniques as FROSt [19] or TIPTOE [20].

In the last few years, another spectrographic technique named amplitude swing (a-swing) was introduced [21] and further developed across diverse scenarios. It scans the relative amplitude between two delayed replicas of the input pulse using a bulk interferometer, based on a rotating multiple-order waveplate (MWP). Since it presents a compact in-line implementation, the system is stable, robust, and easy to align. Furthermore, it is a very versatile technique: it can measure pulses in a broad range of temporal durations and spectral bandwidths [21,22], including the few-cycle regime [23], operating at different spectral regions [24], and can also characterize vector pulses, whose polarization state is time-dependent, from a single trace [25]. Alternatively, the amplitude modulation can be applied using a rotating zero-order waveplate (ZWP) followed by a static MWP, as it is done in the generalizing configuration [26]. In Ref. [27], a unified theory of a-swing was proposed, encompassing said amplitude modulations by means of a propagation matrix, which can be defined for any implementation.

In this paper, we present a single-shot a-swing implementation, maintaining the mentioned advantages of the scanning configurations. We demonstrate this technique, first analytically and numerically (Section 2), and then experimentally (Section 4), by comparing the single-shot a-swing trace (spatially resolved) with the generalizing a-swing trace and their corresponding retrieved pulses.

## 2. Fundamentals of single-shot a-swing

The a-swing technique consists in generating two replicas of the pulse to be measured with a fixed time delay (on the order of the pulse's Fourier limit) between them, while varying their relative amplitude. For each relative amplitude, the spectrum of a nonlinear signal (e.g., SHG) of the interference pulse is recorded, thus obtaining a two-dimensional trace that encodes the amplitude and phase of the input pulse [21,27]. This is achieved with a rotating MWP (conventional configuration [21]), or with a rotating ZWP followed by a static MWP (generalizing set-up [26]), in the scanning implementations. Here, the objective is to replace the rotating scan by encoding the amplitude modulation in a spatial dimension, so that it can be recorded in a single shot, for which we propose the setup presented in Fig. 1. The two replicas with modulated amplitude are created by a pair of uniaxial crystal wedges (oriented with perpendicular optical axes) and a zero-order quarter-wave plate (QWP), whereas the delay is introduced by the following MWP, as detailed next.

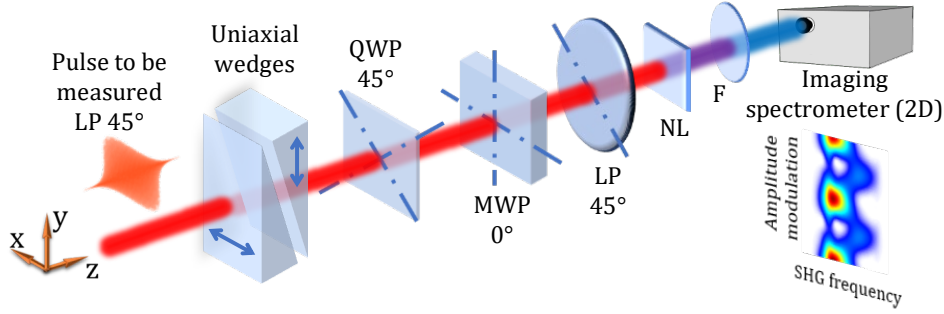

Fig. 1. Single-shot a-swing setup scheme: the input pulse, linearly polarized at 45º, propagates through a pair of birefringent wedges (optical axes horizontal and vertical, indicated by the arrows), a zero-order quarter-wave plate (QWP, fast axis at 45º), a multiple-order waveplate (MWP, fast axis horizontal), and a linear polarizer (LP, transmission axis at 45º). The resulting modulated beam passes through a nonlinear crystal (NL), filtering out the remaining fundamental signal (F), and collecting the spatially mapped SHG signal (i.e., the single-shot a-swing trace) using an imaging spectrometer.

The input pulse —spatially homogeneous in this analysis, propagating along the $z$-axis and linearly polarized at 45º, so its Jones vector is $E(\omega)(1,1)$— passes through a pair of uniaxial wedges, whose optical axes are perpendicular between them, both contained in the incident face and drawing a 45º angle with the polarization direction of the pulse. Let's call $\rho_{x,y}(\omega, y)$ the absolute phases acquired by the horizontal ($x$) and vertical ($y$) components of the input pulse, respectively. Note that these phases depend on the $y$ coordinate, due to the thickness variation of each wedge (see the drawing in Fig. 1). Thus, the pulse after the wedges is:

$$\begin{pmatrix} E_x^w(\omega, \delta) \\ E_y^w(\omega, \delta) \end{pmatrix} = \begin{pmatrix} E(\omega)\, e^{i\,\rho_x(\omega,y)} \\ E(\omega)\, e^{i\,\rho_y(\omega,y)} \end{pmatrix} = e^{i\,\rho_x(\omega,y)} \begin{pmatrix} E(\omega) \\ E(\omega)\, e^{-i\,\delta(\omega,y)} \end{pmatrix} \quad (1)$$

where $\delta(\omega, y) = \rho_x(\omega, y) - \rho_y(\omega, y)$ is the relative dispersion due to the wedges. This means that, for each frequency $\omega$, the beam polarization azimuth is ±45° and the ellipticity depends on $\delta$, which varies linearly along the $y$-axis (Fig. 2), as described below. The phase of each wedge is:

$$\rho_{x,y}(\omega, y) = k_{x,y}(\omega) \cdot d(y) = \frac{\omega}{c}\, n_{x,y}(\omega)[d_0 \mp y \tan \alpha] \quad (2)$$

where the upper and lower signs apply to 1st and 2nd wedge, respectively. Here, $k_{x,y}(\omega)$ represent the wavenumber of the $x$- and $y$- axes; $d(y)$ and $\alpha$ are the thickness and angle of the wedges, respectively, being $y = 0$ the vertical position where both wedges have the same thickness, $d_0$, resulting in zero accumulated relative phase; and $n_{x,y}(\omega)$ are the refractive indices applying to the $x$- and $y$- axes. Thus, the absolute and relative phases acquired during the propagation through both wedges are:

$$\rho_{x,y}(\omega, y) = \frac{\omega}{c}\, (d_0[n_o(\omega) + n_e(\omega)] \pm y \tan \alpha\, [n_o(\omega) - n_e(\omega)])$$
$$\delta(\omega, y) = \rho_x(\omega, y) - \rho_y(\omega, y) = \frac{\omega}{c}\, 2y \tan \alpha\, [n_o(\omega) - n_e(\omega)] \quad (3)$$

with the upper and lower signs corresponding to $\rho_x$ and $\rho_y$, respectively, and where it is replaced $n_x = n_e$ and $n_y = n_o$ in the first wedge, and on the contrary in the second wedge, being $n_o(\omega)$ and $n_e(\omega)$ the ordinary and extraordinary refractive indices of the wedges material. Due to this birefringence, both beams propagate in slightly different directions after the first wedge, while being parallel and subtly sheared after the second wedge, but this effect is negligible for the parameters required in the present application. Note that it depends on the wedge angle and material birefringence, both having low values, e.g., using quartz wedges of 2º, the relative deviation for 800 nm is 0.02º. The birefringence dispersion also implies that the relative phase $\delta$ is slightly different (for moderate spectral bandwidths as in the results presented here) for each wavelength, which anyway can be simply considered in the simulations and retrieval algorithm (as we do here), though it is omitted in the following explanation for clarity.

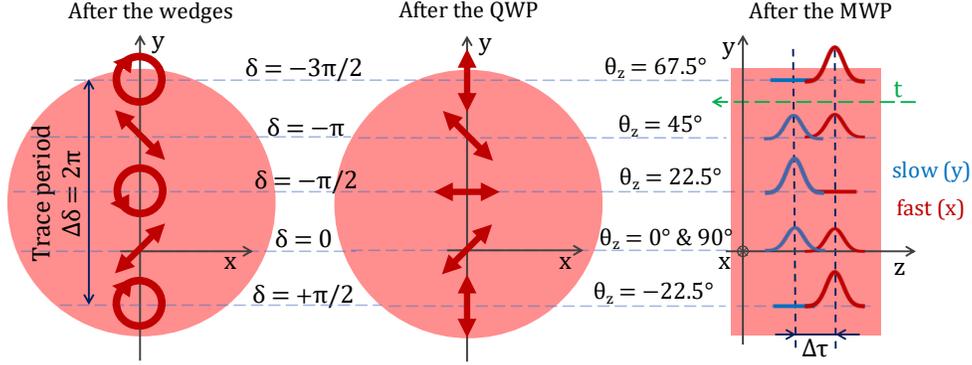

Fig. 2. Amplitude modulation scheme in the single-shot a-swing, considering a positive uniaxial crystal (for a negative crystal, $\delta > 0$ when $y > 0$). After the wedges (left), the polarization ellipse has a ±45º azimuth and a space-varying ellipticity (along the $y$-axis), degenerating in linear and circular polarization when $\delta = \pi m$ and $\delta = \pi/2 + \pi m$, respectively (being $m$ an integer). After the QWP (center), the beam is linearly polarized, and the polarization direction rotates along the $y$-axis. (Right) The MWP introduces a time delay ($\Delta\tau$) between the horizontal and vertical components of the pulse after que QWP, resulting in two delayed replicas of the pulse with spatially varying relative amplitude. When $\delta = \pi m$, both replicas have the same intensity, and when $\delta = \pi/2 + \pi m$, there is only one replica, the same as in the HWP generalizing a-swing when $\theta_z = m \cdot 45°$ and $\theta_z = 22.5° + m \cdot 45°$, respectively.

The QWP, with its fast axis at 45º, transforms the $x$- and $y$- components of the pulse after the wedges into right- and left-handed circular polarization, respectively. Their sum with the same amplitude (as is the case here) gives a linearly polarized beam, whose orientation depends on their phase difference. Thus, after the QWP, the beam is linearly polarized, and its orientation, given by the azimuth, varies gradually along the $y$-axis (Fig. 2) depending on $\delta(\omega, y)$, the phase difference introduced by the wedges between said projections. This space-dependent polarization is similar to the one imparted in other works along the azimuthal coordinate using s-plates for vortex generation [28,29], whereas in this application, it is applied along a longitudinal axis, obtaining the following vector beam:

$$\begin{pmatrix} E_x^{QWP}(\omega,\delta) \\ E_y^{QWP}(\omega,\delta) \end{pmatrix} = E(\omega) \frac{e^{i\,\rho_x(\omega,y)}}{\sqrt{2}} \begin{pmatrix} 1 - i\,e^{-i\,\delta(\omega,y)} \\ -i + e^{-i\,\delta(\omega,y)} \end{pmatrix} = \frac{E(\omega)\sqrt{2}\,e^{i\,\rho_x(\omega,y)}}{e^{i\left(\frac{\delta(\omega,y)}{2}+\frac{\pi}{4}\right)}} \begin{pmatrix} \cos\left[\frac{\delta(\omega,y)}{2} + \frac{\pi}{4}\right] \\ \sin\left[\frac{\delta(\omega,y)}{2} + \frac{\pi}{4}\right] \end{pmatrix} \quad (4)$$

This way, the relative amplitude between the horizontal and vertical projections is modulated along the $y$-axis. These projections are temporally delayed by the following MWP (Fig. 2), where the pulse acquires the phases $\rho_f(\omega)$ and $\rho_s(\omega)$ in its fast (corresponding to $x$-axis in this case) and slow ($y$-) axes, respectively:

$$\begin{pmatrix} E_x^{MWP}(\omega,\delta) \\ E_y^{MWP}(\omega,\delta) \end{pmatrix} = E(\omega)\sqrt{2}\,e^{i\,\rho_x(\omega,y)}\,e^{-i\left(\frac{\delta(\omega,y)}{2}+\frac{\pi}{4}\right)} \begin{pmatrix} e^{i\,\rho_f(\omega)}\cos\left[\frac{\delta(\omega,y)}{2} + \frac{\pi}{4}\right] \\ e^{i\,\rho_s(\omega)}\sin\left[\frac{\delta(\omega,y)}{2} + \frac{\pi}{4}\right] \end{pmatrix} \quad (5)$$

These two delayed replicas with space-dependent relative amplitude are projected into the transmission axis of the following linear polarizer (LP), oriented at 45º, so the interference pulse is:

$$E_{45°}^{LP}(\omega, y) = \frac{E(\omega)\, e^{i\,\rho_x(\omega,y)}}{e^{i\left(\frac{\delta(\omega,y)}{2}+\frac{\pi}{4}\right)}} \left( e^{i\,\rho_f(\omega)} \cos\left[\frac{\delta(\omega,y)}{2}+\frac{\pi}{4}\right] + e^{i\,\rho_s(\omega)} \sin\left[\frac{\delta(\omega,y)}{2}+\frac{\pi}{4}\right] \right) \quad (6)$$

and the SHG spectrum of this pulse constitutes the a-swing trace:

$$S_{45°}^{SHG}(\omega, y) = \left|E_{45°}^{SHG}(\omega, y)\right|^2 = \left|\int \left(\int E_{45°}^{LP}(\omega', y)\, e^{i\omega' t}\, d\omega'\right)^2 e^{-i\omega t}\, dt\right|^2 \quad (7)$$

This spatially resolved trace is equivalent to the trace obtained in the generalizing a-swing configuration [26] with a rotating zero-order half-wave plate (HWP) (Fig. 3), in which the amplitude modulation of the delayed replicas is described by sine and cosine functions. Here, the rotating HWP is substituted by a pair of uniaxial wedges and a QWP. The amplitude modulation, imparted by the rotating angle in the scanning setup, is translated here into the vertical spatial dimension, allowing to measure the a-swing trace in a single acquisition using an imaging spectrometer. This spatially mapped trace is ideally repeated each $2\pi$ interval of $\delta(\omega, y)$ and the existing equivalence between the scanning and single-shot axes is $2\theta_Z \leftrightarrow -\delta/2$, being $\theta_Z$ the fast axis orientation of the rotating HWP in the generalizing a-swing [26]. The corresponding equivalence with the vertical axis is given by $\delta(\omega, y)$, which depends on the angle and birefringence of the wedges (Eq. 3) and can be experimentally calibrated as detailed in Section 3. The QWP can be replaced by any other ZWP, obtaining a different amplitude and phase modulation (as in the scanning version [26]), which could also be used to measure the pulse provided that the system is calibrated. To reconstruct the pulse, we have adapted a ptychographic iterative engine (PIE) algorithm [27] (explained in the following section), which we have initially tested with simulated traces (an example is shown in Fig. 3).

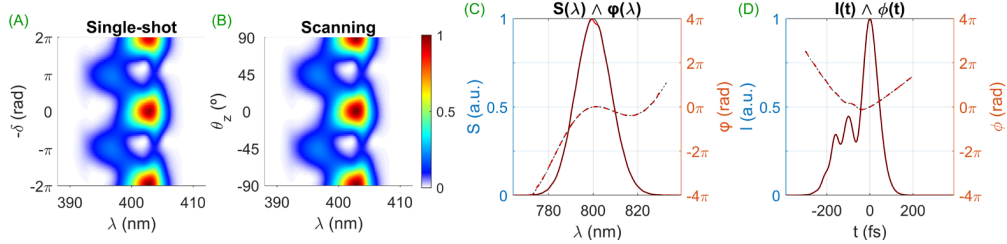

Fig. 3. Simulated single-shot (A) and HWP generalizing (B) a-swing traces (the retrieved traces are equivalent to the simulated ones). Simulated (black) and retrieved (red) intensity (solid) and phase (dashed) in the spectral (C) and temporal (D) domains. The pulse is defined by a gaussian spectrum centered at 800 nm, Fourier-limited pulse duration of 50 fs and an oscillating spectral phase. The MWP used for the simulation introduces a 50-fs delay and a $0.7\pi$ rad phase retardation for 800 nm. The wedges material is quartz, their central thickness is 1 mm and their angle is 2°. Note that, for comparison purposes, the single-shot trace vertical axis is $-\delta$, while the scanning angle is plotted from $-90°$ to $90°$.

## 3. Retrieval algorithm and wedges calibration

The single-shot a-swing trace (Eq. 5 and 6) depends on the input pulse $E(\omega)$, the MWP spectral phases $\rho_{f,s}(\omega)$, and the wedges absolute and relative phases, $\rho_{x,y}(\omega, y)$ and $\delta(\omega, y)$, respectively. Both the MWP and the pair of wedges can be experimentally calibrated, so the only unknown is the input pulse.

Regarding the calibration of the elements, in the case of the MWP, it is characterized by its delay and phase retardation. These parameters can be obtained from the MWP thickness, precisely calculated by using in-line SI [30], and the ordinary and extraordinary refractive indices, modeled by the corresponding Sellmeier equations [31]. The absolute phase introduced by the wedges is obtained as the MWP dispersion, from the thickness and refractive indices. Regarding the relative phase $\delta$, the orientation of the wedges and their optical axes determine its sign (in the implementation shown in Fig. 1, $\delta > 0$ for $y > 0$ for negative uniaxial crystals as calcite, and on the contrary for positive crystals as quartz). The material, optical axes orientation, angle, and thickness are typically provided by the manufacturer. In this case, we only need to determine the vertical position corresponding to $\delta = 0$, which can be easily known by rotating the last LP in the setup (Fig. 1). From Eq. 5, it is deduced that the beam is horizontally and vertically polarized when $\delta = -\pi/4 + 2\pi m$ and $\delta = +\pi/4 + 2\pi m$, respectively. Thus, rotating at 0º the last LP of the setup (Fig. 1), the spatially resolved spectrum (both fundamental and SHG) presents maxima and minima when $\delta = -\pi/4 + 2\pi m$ and $\delta = +\pi/4 + 2\pi m$, respectively (and on the contrary for the LP at 90º). Another option is to do this calibration without the MWP, so the previous reasoning is valid for any LP orientation, so we could even determine the sign of $\delta$ (in case it was unknown). Furthermore, the distance between maxima or minima works as a cross-check measurement by comparing it with the theoretical result, and can be used, e.g., to calibrate the pixel size of the imaging spectrometer. An alternative approach to fully characterize the wedges involves spatially resolved in-line SI [32], using the same setup (Fig. 1), increasing the delay by replacing the a-swing MWP or adding another plate, and collecting the interferometric fundamental spectrum. From this measurement, we directly obtain the function $\delta(\omega, y)$ using Fourier analysis. The SI approach is more precise, thus it is suitable for the first calibration of the wedges, while the LP rotation method is simpler, so it can be recommended for the subsequent measurements. This latter calibration does not have to be repeated unless the wedges are moved.

With respect to the pulse reconstruction, the input pulse is defined as $E(\omega) = A(\omega)\,e^{i\,\varphi(\omega)} = \sqrt{S(\omega)}\,e^{i\,\varphi(\omega)}$, being $A(\omega)$, $S(\omega)$, and $\varphi(\omega)$ the spectral amplitude, intensity, and phase, respectively. To reconstruct the pulse from the spatially resolved a-swing trace, we have used a PIE algorithm. Such algorithm was originally applied to diffractive imaging [33], and later adapted to FROG [34], FROSt [6], d-scan [35] and a-swing [27] reconstructions. As we will describe, this algorithm does not consider the relative amplitude between the SHG spectra for different $\theta$ or $\delta$ values, i.e., between different trace slices, since it optimizes them individually at each iteration. In the single-shot implementation, ideally the beam is homogeneous. Even if the pulse is constant from point to point, the intensity profile can spatially vary thus modulating the trace in the $y$ dimension, so the PIE algorithm suits perfectly since this modulation would not affect the pulse reconstruction. On the contrary, other algorithms such as Levenberg-Marquardt or differential evolution, which perform well in scanning traces retrievals [21–26], are less adequate to reconstruct single-shot measurements since they optimize the whole trace at each iteration (i.e., all the SHG spectra simultaneously). Other strategies could be used to deal with inhomogeneous beam profiles, for example, generating a flap-top beam, rescaling the experimental trace from the measured beam profile, or normalizing each SHG spectrum of the experimental and reconstructed traces before comparing them (note that this results in a loss of information that may hinder the reconstruction). Anyway, here our choice of the PIE algorithm is simpler and more appropriate since we do not need spatial filters or shapers, additional beam profile measurement, nor posterior trace processing. Furthermore, PIE is faster than the other mentioned strategies and can extract not only the phase but also the amplitude of the field [27]. Briefly, this algorithm works as follows, being $k$ the loop iterator through the $y$- (or $\delta$-) axis ($\theta$ in the scanning versions). A pulse with a flat spectral amplitude (or the measured one) and a random or flat spectral phase serves as an initial guess. We calculate the propagation matrix $\mathbf{H}(\omega, y)$ of the single-shot a-swing shaping system, i.e., the wedges, QWP and MWP:

$$\mathbf{H}(\omega, y) = \frac{e^{i\,\rho_x(\omega,y)}}{\sqrt{2}} \begin{pmatrix} e^{i\,\rho_f(\omega)} & -i\,e^{i\,\rho_f(\omega)}\,e^{-i\,\delta(\omega,y)} \\ -i\,e^{i\,\rho_s(\omega)} & e^{i\,\rho_s(\omega)}\,e^{-i\,\delta(\omega,y)} \end{pmatrix} \qquad (8)$$

which is then applied to the guess input (unknown) pulse, obtaining the forward propagated pulse, the vector pulse before the LP at 45º just before the nonlinear crystal. As detailed in [27], this LP makes the propagation non-invertible. To overcome it, we update the projection at +45º (the interference pulse) as explained below, while saving the calculated −45º component without update for later computation. Using the inverse Fourier-transform, we get this guess interference pulse in the time domain $E_k(t)$ for a random $k$, i.e., a random $\delta$. Then, we calculate its corresponding SHG electric field in time domain, $\psi_k(t)$, which is then amended in the spectral domain by replacing its amplitude by the measured nonlinear spectrum $\sqrt{S_k^{\text{SHG}}(\omega)}$ (Eq. 7). Going back to time domain, we get a new SHG signal, $\psi'_k(t)$. In the next step, the interference pulse is renewed by the update equation:

$$E'_k(t) = E_k(t) + \beta_k \frac{E_k(t)^*}{|E_k(t)|^2 + \alpha\,|E_k(t)|^2_{\max}} [\psi'_k(t) - \psi_k(t)] \qquad (9)$$

where $\beta_k$, randomly chosen between 0.1 and 0.5, controls the update strength, and $\alpha$, set to 0.15, avoids dividing by zero. By Fourier-transforming it to the frequency domain, we get the updated interference pulse. Together with the unchanged projection at −45º, this yields the updated forward pulse. By backpropagating this pulse, multiplying it by **H⁻¹**, we get the unknown pulse that is used as a new guess. At this stage, if the fundamental spectrum has been measured (optional), it can be used to replace the reconstructed pulse amplitude. Starting from the new guess pulse, this process is repeated for other randomly picked $k$ slice of the trace, and so on until the entire trace is covered (in our case, 91 $k$ values), thus completing one main iteration. This main loop is repeated until convergence, needing approximately 25 full iterations, which are done in ~20 s.

### 4. Experimental demonstration

*4.1 Experimental implementation*

The laser is a Ti:sapphire chirped pulse amplified (CPA) system (Spitfire ACE, Spectra Physics). The central wavelength is ~800 nm, the Fourier-limited pulse duration is ~60 fs, and the repetition rate is 5 kHz. Before the a-swing setup (Fig. 1), the pulses emitted by this system propagate through a 70-cm length hollow-core fiber (HCF) at two different conditions: vacuum (~10⁻³ mbar), and small air pressure (~450 mbar). In the second case, nonlinear propagation slightly modifies the fundamental spectrum. The HCF is introduced to test the system with different fundamental spectra, while increasing the quality of the beam, which results in a benefit for the presented (and for any) spatially resolved single-shot technique. Regarding the beam size, resolution and aperture of the imaging spectrometer, and wedges dimensions, they must be designed to get at least one full trace in the sensor. This means that the clear aperture of the system and the beam size must be larger than a trace period. Furthermore, the spatial resolution of the imaging spectrometer must allow to measure enough spectra within a trace period (e.g., at least 90 values as in the scanning version).

We used the experimental scheme shown in Fig. 1. In these experiments, the pair of wedges with crossed optical axes is replaced by a pair of wedges with their optical axes parallel between them (horizontal), and a HWP oriented at 45º between the wedges to make the system equivalent. By changing the phase retardation of the ZWP placed between or after the wedges, the amplitude and phase modulation along the spatial coordinate could be modified, resulting in different a-swing traces that would encode the pulse information alike [26]. Here, we used quartz wedges (2º angle, 40-mm height, 0.5-mm minimum thickness and 1.75-mm maximum thickness, from Altechna). As explained in Section 2, for an input pulse linearly polarized at 45º, the wedges introduce a phase difference $\delta(\omega, y)$ (Eq. 3) between the horizontal and vertical components, which after the QWP at 45º creates a gradually rotating polarization along the *y*-axis (Fig. 2). The relative phase $\delta(\omega, y)$ varies along the *y*-axis at a rate of $2\pi$ rad per ~1.28 mm for 800 nm, which defines the trace period, in which the final linear polarization orientation rotates by 180º. The horizontal and vertical components of the rotating linear polarization (amplitude varying along the *y*-axis) are delayed by a MWP (quartz, ~2-mm thickness), which introduces a 65-fs delay and a $0.34\pi$ phase retardation (for 800 nm). As in the scanning implementations, the choice of MWP delay is related to the Fourier-limit pulse duration, accepting a range defined by a factor of 3 [21]. The latter LP selects the intermediate projection, which then produces SHG in a type-I BBO crystal (20-µm thickness to ensure flat phase matching efficiency). The remnant fundamental light is removed with a bandpass filter (F; BG-39, Schott), recording the spatially mapped SHG spectrum with an imaging spectrometer (Shamrock SR-303i-B, Andor Technology Ltd). Optionally, a cylindrical lens can focus the SHG signal on the spectrometer (vertical) slit for better SHG signal collection, as we did in the second single-shot trace measurement. We have measured the fundamental spectrum with a conventional spectrometer (AvaSpec 2048-USB1, Avantes Inc.).

To validate the single-shot retrievals, we performed scanning HWP generalizing a-swing measurements to compare both the experimental traces (they are equivalent) and the corresponding reconstructed traces and pulses. The experimental procedure to switch between the single-shot and scanning setups is straightforward, since most of the elements are the same: we just replace the QWP by a rotating HWP (this change is just for comparison purposes, but any same ZWP could be used for both setups), the imaging spectrometer by a conventional spectrometer (the imaging could be used too), and set to 0º the input pulse polarization direction (so the wedges do not modulate the beam, we remove the HWP in between or set it to 0º) and the MWP (45º) and LP (0º) axes orientation accordingly [26]. The pulses propagate through identical elements in both approaches, ensuring consistent dispersion that is accounted for in the algorithm to retrieve the pulse before the measurement devices, requiring no additional adjustments for comparison.

### 4.2 Experimental results and discussion

We present the comparison between the single-shot and scanning generalizing a-swing measurements of two pulses that propagate through an HCF (Fig. 4). First, the HCF is in vacuum, thus the propagation is linear, without spectrum modification (Fig. 4.1). Then the HCF is filled with small air pressure, so the spectrum is slightly modified due to nonlinear propagation (Fig. 4.2). In both cases, the measured spatially mapped and scanning traces (Fig. 4A and 4C) are equivalent, as expected. We can observe small differences between the measured and retrieved single-shot traces, which may be attributed to the beam inhomogeneity and non-ideal achromatic half-wave operation of the HWP between the wedges. The corresponding retrieved pulses from the single-shot and scanning a-swing, using the PIE algorithm (Fig. 4E and 4F), also present a very good agreement, thus validating the single-shot implementation.

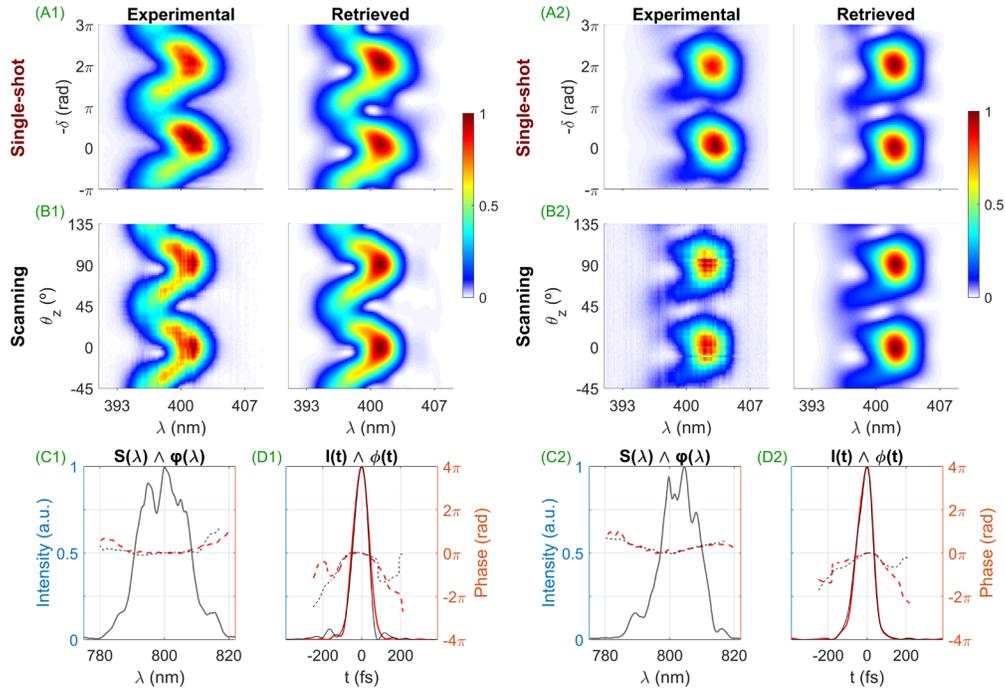

Fig. 4. Measurement of pulses propatating through a HCF in vacuum (1) and air at ~450 mbar (2). Experimental and retrieved single-shot (A) and HWP generalizing (B) traces; corresponding retrieved pulses in the spectral (C) and temporal (D) domains. Solid: measured spectrum (grey) and retrieved temporal intensity; dashed: retrieved spectral and temporal phases. Red: single-shot; black: HWP generalizing. Note that, for comparison purposes, the single-shot trace vertical axis is $-\delta$, while the scanning trace is plotted from $-45°$ to $135°$.

## 5. Conclusions

We have demonstrated a single-shot implementation of the temporal characterization technique amplitude swing, maintaining its benefits: compact in-line implementation, robustness, and versatility. The amplitude swing concept, implemented in scanning previous versions using a rotating multiple-order waveplate or a rotating zero-order waveplate combined with a static multiple-order waveplate, is realized here in a single-shot architecture by a pair of uniaxial wedges. Combined with a zero-order quarter-wave plate, this setup generates a spatially varying relative amplitude and phase between two pulse replicas, which are then temporally delayed by a multiple-order waveplate.

We first theoretically studied the proposed setup, showing that the resultant spatially mapped trace corresponds to the generalizing amplitude swing trace using a rotating zero-order half-wave plate. Then, we experimentally validated the single-shot setup by comparing the measured single-shot and scanning traces, and their corresponding retrieved pulses, which highly agree. For reconstructing the pulse from the spatially mapped trace, measured using an imaging spectrometer, we have adapted a ptychographic iterative engine (PIE) algorithm, recently proposed for scanning trace retrievals.

In sum, we present a single-shot method for measuring ultrashort laser pulses, with a simple setup, which can cover a wide range of spectral regions, bandwidths (including the few-cycle regime), and chirps, while being resilient against noise and clipping. This way, we address the temporal characterization of single pulses, required when handling pulse-shape instabilities, common in high-power or low repetition rate lasers, even in single pulse operation, and in unstable experiments.

**Funding.** Ministerio de Ciencia e Innovación (PID2020-119818GB-I00, PID2023-149836NB-I00); Consejería de Educación, Junta de Castilla y León (SA108P24); European Regional Development Fund.

**Acknowledgements.** C. Barbero acknowledges the support from Junta de Castilla y León and Fondo Social Europeo Plus through their Ph. D. grant program.

**Disclosures.** B. Alonso and Í. J. Sola have patent # WO/2021/123481 pending to University of Salamanca.

**Data availability.** Data underlying the results presented in this paper are not publicly available at this time but may be obtained from the authors upon reasonable request.